\documentclass[iop,revtex4]{emulateapj-rtx4}
\usepackage{natbib}
\usepackage{lscape}
\bibpunct{(}{)}{;}{a}{}{,}
\usepackage{graphicx,epsfig}%,emulateapj5}
\newcommand{\eten}[1]{\mbox{$10^{#1}$}}% power of ten

\newcommand{\kms}{\mbox{km s$^{-1}$}}% km/s

\newcommand{\lsun}{\mbox{L$_\odot$}}% Lsun
\newcommand{\msun}{\mbox{M$_\odot$}}% Msun

\newcommand{\tmbdv}{\mbox{$T_{\rm MB}$ d$V$}}

\newcommand{\imb}{\mbox{$I_{\rm MB}$}}

\newcommand{\lbol}{\mbox{$L_{bol}$}} % bolometric luminosity
\newcommand{\tbol}{\mbox{$T_{bol}$}} % bolometric temperature

\newcommand{\mean}[1]{\mbox{$\langle#1\rangle$}} %generic mean for defined qu.
\newcommand{\av}{\mbox{$A_V$}} % Visual Extinction
\newcommand{\coo}{$^{13}$CO}

\newcommand{\hcop}{HCO$^+$}

\newcommand{\etamb}{\mbox {$\eta_{\rm MB}$}}

\newcommand{\jj}[2]{\mbox{$J = #1\rightarrow#2$}}
\newcommand{\minf}{\mbox{$\dot M_{inf}$}}
\newcommand{\mstar}{\mbox{M$_{\star}$}}

\newcommand{\kkms}{\mbox{K\ \kms}}

\shorttitle {The Real Solar Neighborhood Protostars}
\shortauthors{Heiderman \& Evans}

\begin{document}

\title{The Gould Belt `MISFITS' Survey: The Real Solar Neighborhood Protostars}

\author{Amanda Heiderman\altaffilmark{1,2,3,4}}
\author{Neal J. Evans II\altaffilmark{5}}
\altaffiltext{1}{Department of Astronomy, University of Virginia, P.O. Box 400325, Charlottesville, VA 22904, USA; heiderman@virginia.edu}
\altaffiltext{2}{National Radio Astronomy Observatory, 520 Edgemont Road, Charlottesville, VA 22903, USA}
\altaffiltext{3}{Max-Planck-Institut f\"{u}r Astronomie, K\"{o}nigstuhl 17, D-69117 Heidelberg, Germany}
\altaffiltext{4}{NSF Astronomy and Astrophysics Postdoctoral Fellow}
%\author{Neal J. Evans II}
\altaffiltext{5}{Department of Astronomy, The University of Texas at Austin, 2515 Speedway, Stop C1400, Austin, Texas 78712-1205, USA; nje@astro.as.utexas.edu} 

%\email{nje@astro.as.utexas.edu}

% 012015 ALH
% 012915 NJE
% 020415 NJE

\begin{abstract}\label{abstract}

We present an \hcop\ \jj32\ survey of Class 0+I and Flat SED young stellar objects (YSOs) found in the Gould Belt clouds by surveys with $Spitzer$. 
Our goal is to provide a uniform Stage 0+I source indicator for these embedded protostar candidates. We made single point \hcop\ \jj32\  measurements toward the source positions at the CSO and APEX of 546 YSOs (89\% of the Class 0+I $+$ Flat SED sample).  Using the criteria from \citet{2009A&A...498..167V},  we classify sources as Stage 0+I or bona fide protostars and find that 84\% of detected sources meet the criteria.
We recommend a timescale for the evolution of Stage 0+I (embedded protostars)
of 0.54 Myr.  We find significant correlations of \hcop\ 
integrated intensity with $\alpha$ and \tbol\ but not with \lbol.  
The detection fraction increases
smoothly as a function of $\alpha$ and \lbol, while decreasing smoothly
with \tbol.
Using the Stage 0+I sources tightens the relation between protostars and 
high extinction regions of the cloud; 89\% of Stage I sources 
lie in regions with \av\ $> 8$ mag.   Class 0+I and Flat SED YSOs that are not detected in \hcop\ have, on average, a factor of $\sim2$ higher \tbol\ and a factor of $\sim$5 lower \lbol\ than YSOs with \hcop\ detections. 
We find less YSO contamination, defined as the number of undetected YSOs 
divided by the total number surveyed, for sources with $\tbol \lesssim 600$ K 
and $\lbol \gtrsim 1$ \lsun. The contamination percentage is $>90$\% 
at \av$<$4 mag and decreases as \av\ increases.
\end{abstract}

\keywords{ISM: clouds; dust; extinction; galaxies: ISM; galaxies: star formation; infrared: stars; stars: formation}

% intro.tex
% 020714 NJE
% 082014 NJE
% 102214 NJE
% 111014 ALH 
% 111914 NJE
% 012015 ALH
% 012915 NJE
% 020515 NJE
% 022115 NJE fixed Class/Stage issues raised by referee and added bf

\section{Introduction}\label{intro}
The evolution of dense molecular cores into stars has been characterized
by observational changes in their spectral energy distributions (SEDs)
since the seminal suggestions by 
%Lada Wilking, original class system
\citet{1984ApJ...287..610L}
and
%Adams, Lada, and Shu 1987
\citet{1987ApJ...312..788A}.
In this picture the changes in the SED,  captured either by the
spectral index from 2 to 25 \micron\ in the original Class system,
or the bolometric temperature
% Chen et al.
\citep{1995ApJ...445..377C},
track the movement of matter from core to disk to star.
%ALS 
\citet{1987ApJ...312..788A}
connected the observational measures of the SED (Classes) to physical
configurations (Stages). 
The stage referred to as I in the original work included
all the time that the forming star was surrounded by an envelope,
a phase also referred to as a protostar, or the embedded phase.

%Andre et al. 1993
\citet{1993ApJ...406..122A}
sub-divided Stage I into 0 and I, with Stage 0 sources having more mass
in the envelope than in the star plus disk.  Stage I sources had
more mass in the star plus disk than in the envelope, but still some
envelope material. We have no way to distinguish these, so
we lump these together in this paper, referring to them as Stage 0$+$I.
Thus, a Stage 0$+$I source consists of a star and
disk embedded in a dense, infalling envelope, 
while a Stage II source contains only a star and disk.

The exact relationship between observations and
physical configuration required further definition.
%Robitaille et al.
\citet{2006ApJS..167..256R}
suggested defining Stage 0$+$I sources as those with 
substantial mass infall rates (\minf) compared to stellar
masses (\mstar), requiring $\minf /\mstar > \eten{-6}$ yr$^{-1}$.
However, these two quantities are difficult to determine from
observations, and \minf\ is not related monotonically to 
evolutionary Stage in models of episodic accretion 
%(e.g., Dunham et al. 2010, 2012
\citep{2010ApJ...710..470D,2012ApJ...747...52D}.
%Crapsi et al
\citet{2008A&A...486..245C} found that the usual evolutionary tracers,
like the spectral index in the near-infrared to mid-infrared or the bolometric
temperature (\tbol) could be misleading for sources with large inclination
angles, but that 0.2 \msun\ in the envelope was a reasonable 
indicator to select Stage 0$+$I sources.

%Van Kempen et al. 
\citet{2009A&A...498..167V} showed that detection of the \hcop\
\jj43\ line was a good indicator of a Stage 0$+$I source. In particular
they suggested that an integrated intensity,
\imb$ \equiv \int$ \tmbdv,  of 0.4 \kkms\
is a good metric for a Stage 0$+$I source, because lower values of 
integrated intensity could arise if only a disk were present. 
To further discriminate against disk emission and unrelated
emission from sources nearby in the sky, 
they also suggested that the
\hcop\ emission should be extended, but peaked on the source, or
that submillimeter dust continuum emission be extended but peaked on the source.

The c2d 
%(Evans et al.) 
\citep{2009ApJS..181..321E}
and Gould Belt (Dunham, in prep.) 
projects together surveyed
essentially all nearby ($d < 500$ pc) molecular clouds, except for Taurus and
Orion. They used a common classification scheme for YSOs based on the spectral
index between 2 and 24 \micron, using whatever photometry was available between
those wavelengths. The spectral index is defined by
\begin{equation}
\alpha = \frac{d\log(\lambda S(\lambda))}{d\log(\lambda)},
\end{equation}
where $S(\lambda)$ is the spectral flux density at wavelength $\lambda$.

Building on work by 
%Greene et al.
\citet{1994ApJ...434..614G},
\citet{2009ApJS..181..321E} defined
the divisions between classes as follows:

\begin{description}
\item[0$+$I] $0.3 \leq \alpha $;
\item[Flat SED] $-0.3 \leq \alpha <0.3$;
\item[II] $-1.6 \leq \alpha < -0.3$;
\item[III] $\alpha < -1.6$.
\end{description}
The Class 0$+$I sources are widely associated with Stage 0$+$I, while Class II
sources are associated with Stage II (star and disk dominated objects), but
the status of Flat SED sources has been unclear.

When using $\alpha$ alone, there is no distinction between Class 0 and Class I
sources, so both are included when we use Class 0$+$I, 
and Stage 0$+$I encompasses
the entire phase with a star/disk and surrounding envelope with $M > 0.2$ \msun.

Our goal is to provide a uniform indicator of whether Class 0$+$I 
and Flat SED sources in the survey of the Gould Belt clouds
are likely to be Stage 0$+$I sources. 
Because there were 
over 500 Class 0$+$I and Flat SED sources in the original sample, mapping of each
source was impractical, so we made single point measurements toward
the positions of each YSO instead. We also used the \jj32\ transition
of \hcop, as the atmospheric conditions required are less stringent
than for the \jj43\ line. We describe here the results of 
the MIsidentified YSOs from SED FITS (MISFITS) Survey.

%obs.tex
% 090913 initialized by NJE
% 062414 NJE edits
% 082014 NJE edits
% 102214 NJE
%111314 ALH
% 012915 NJE
% 021515 NJE
% 022115 NJE fixed southern sources, eff. ref,  and added bf

\section{Observations and Spectral Reduction}\label{obs}

Observations of the \hcop\ \jj32\ transition were obtained with the
heterodyne receiver at the Caltech Submillimeter 
Observatory\footnote{The Caltech Submillimeter Observatory 
was operated by the California Institute of Technology, until 2013
April 1  under cooperative agreement with the National Science Foundation 
(AST-0838261)} 
during runs in 2009 (June and December), 2010 (July and December), 2011 (July),
and 2012 (June and September). The main beam efficiencies (\etamb) 
were determined on each run by observations of planets. 
The characteristics of the equipment are
indicated in Table \ref{telescopedat}, including the pointing
uncertainties and the main beam efficiency.
The pointing uncertainties are the standard deviation of all
pointing measurements through the run. Since the pointing offsets
were continually corrected, the actual pointing errors on a given
source are less than the values in the table.
The values of \etamb\ were determined by observations of planets,
generally Jupiter or Saturn. They were relatively constant over 
the various runs with $\mean{\etamb} = 0.67\pm0.03$, 
except for the  2012 June run, when suitable planetary calibrators were not available.
We have used  $\etamb = 0.67$ for all runs, including
the 2012, June run.
%[NJE: it would be good to characterize the mean rms of the CSO obs.]
% Not worth the time it would take

The most southerly clouds, Musca and Chamaeleon, were observed with the
APEX\footnote{APEX is a collaboration between the Max-Planck-Institut 
f\"ur Radioastronomie, the European Southern Observatory, 
and the Onsala Space Observatory.} telescope in 2011, October.  The APEX-1
receiver and the XFFTS spectrometer were used to obtain a
spectral resolution of about $0.085$ \kms.
The forward efficiency of the telescope is $0.95$ and the beam efficiency is $0.75$
(values taken from the APEX website)
corresponding to $\etamb = 0.75/0.95 = 0.79$.
Weather conditions were variable, so integration time was adjusted
to achieve an rms noise of 0.08 K.

The standard data reduction process consisted of averaging multiple
observations when relevant,
box-car averaging to an effective spectral resolution of 0.2 \kms\
for CSO data, $0.17$ \kms\ for APEX data,
fitting a baseline (usually first-order, but occasionally higher order),
and fitting a Gaussian if a line was evident. Because the main
goal was to find out if a detectable line was present, the fitting
process was generally not carefully tuned for optimum parameter 
estimation. In several clouds, multiple velocity components
were present; these were sometimes separable and sometimes too
blended to resolve. When separable, two Gaussians were fitted
and both results are given (see Table \ref{misfitsdat} in Appendix).

When no line was convincingly detected, an upper limit was obtained by
calculating the integrated intensity over a velocity range where lines
were detected in that cloud and the uncertainty in that value.
The upper
limit was taken to be the absolute value of the area plus twice the
uncertainty. Some of the higher upper limits reflect emission that
was systematically high over the  range of relevant velocities 
but did not produce a convincing line profile.

% Table 1  1/16/14 NJE
% 062414 changed efficiency for June 2012 run to mean of others
\begin{deluxetable}{lcc}
%\tabletypesize{\scriptsize}
\tablecaption{Observing Parameters 
\label{telescopedat}}
\tablewidth{0pt}
\tablehead{
\colhead{Run} &
\colhead{Pointing Uncertainty}& 
\colhead{Efficiency} \\%& 
%\colhead{Notes} \\
\colhead{-}  & 
\colhead{(arcsec)} & 
\colhead{-} \\%& 
%\colhead{-} \\
\colhead{(1)} &
\colhead{(2)} &
\colhead{(3)} \\%&
%\colhead{(4)} \\
 }
\startdata
2009 June& 5\farcs9  & 0.67  \\
2009 December& 6\farcs7  & 0.63\\
2010 July & 6\farcs7  & 0.67 \\
2010 December& 4\farcs9	& 0.64 \\
2011 July& 7\farcs8  & 0.66 \\
2011 December& 8\farcs0  & 0.71 \\
2012 June& 10\farcs2 & 0.67\tablenotemark{a} \\
2012 September& 7\farcs6  & 0.69 \\
\enddata
\tablenotetext{a}{ Value set to the mean of all other observing runs because no
suitable calibrator was available.
}
%\end{deluxetable*}
\end{deluxetable}

% results.tex
% NJE 020714
% NJE 041414
% NJE 042814
% NJE 062414 Split into results and analysis
% NJE 070314 added stuff on Class II
% NJE 082014
% NJE 102214
% NJE 111914
% NJE 012915
% NJE 020415

\section{Results}\label{results}

\begin{figure}
\center
\includegraphics[scale=0.5, angle=0]{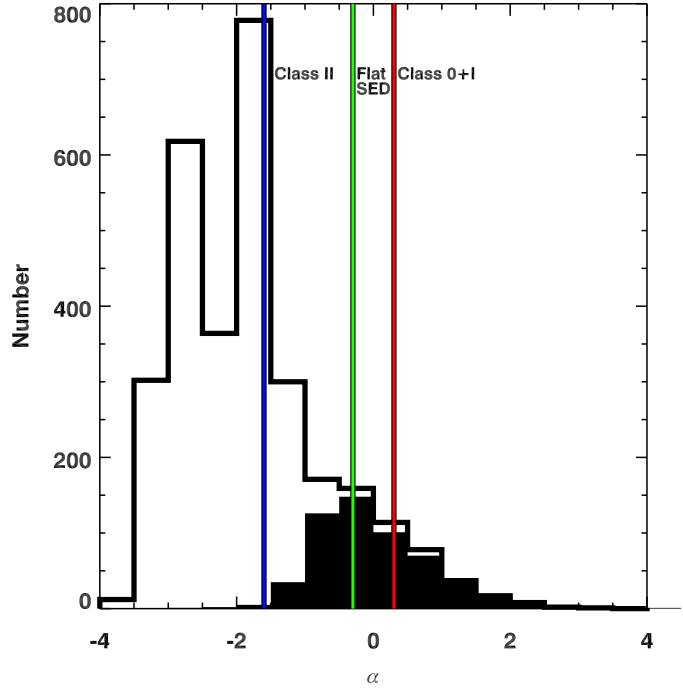}
\caption{
Distribution of $\alpha$ for the full catalog of YSOs (black line) 
and for the observed sample (filled). 
The vertical colored lines mark the boundaries between 
Class II and Class III ($\alpha = -1.6$), Class II and Flat (
$\alpha = -0.3$), and Flat and Class 0$+$I ($\alpha = +0.3$). 
}
\label{hadist}
\end{figure}

For this survey, we compare observations to the table of YSOs
from the work of
% M. Dunham 2013
\citet{2013AJ....145...94D}. 
The full list of sources with \hcop\ observations is given in 
Table \ref{misfitsdat} in the Appendix.
The \hcop\ observations were obtained at positions of sources 
identified in earlier versions of the catalogs. Some of these sources
are no longer accepted as YSOs, a few other sources have been added,
and there are some positional discrepancies. To match observations with
sources in the latest catalogs, we searched for \hcop\ observations
within 14\arcsec\ (half the CSO beam) of the source. However, only 4 sources
lie more than 6\arcsec\ from the pointing position, with the vast majority
still exactly at the pointing center. In a few cases,
more that one observation was found within 14\arcsec. 
The line measurements were similar,
so we used the observations from the nearest position to the infrared 
source. In addition, some sources have changed classifications since the
original catalog was made, and we use the new classifications. 
As a result, we have some sources that are now classified as Class II 
according to the spectral index. We include results on these, but we
caution that they are not representative of most Class II sources because
they are relatively close to the boundary between Class II and Flat sources.
Figure \ref{hadist} shows the distribution of spectral index for the
observed sample and for the full catalog of YSOs. Our observed sample
becomes very incomplete for $\alpha < -0.3$.

The results of the observations are listed in Table \ref{misfitsdat}
in the Appendix.
The summary of results by cloud is in Table \ref{tbl-2}.  
For each cloud with sufficient detections, a composite spectrum
was constructed from all the detected sources. 
From these average spectra, a characteristic velocity
and linewidth were added to Table \ref{tbl-2}. When two velocity
components could be separated in the composite spectrum, values
are given for each.

% analysis.tex
% NJE 062414 Split out from results.tex
% 070214 edits and additions
% 082114 NJE edits
% 102214 NJE
% 102914 ALH edits
% 111914 NJE edits
% 011815 ALH edits
% 012915 NJE
% 020515 NJE
% 022115 NJE fixed Class0/I re Tbol and abs(r) issue, and bias,  and added bf
% 022815 NJE minor edits

\section{Analysis}\label{analysis}

Taking all clouds into account, there are 326 Class 0$+$I sources,
209 Flat SED sources, and 1243 Class II sources in the tables of YSOs from the work
of \citet{2013AJ....145...94D}. Because our survey was based on
earlier catalogs (\S \ref{results}), our coverage is slightly incomplete. 
We observed 288 Class 0$+$I sources (88\% of the full sample), 188 Flat SED sources
(90\% of the full sample), and 70 Class II sources (6\% of the full sample). 
Our statistics will be based on that subsample.

\subsection{Results Based on Any Detection}\label{any}

For a first estimate, we consider all detections, regardless of strength,
within the sample of observed sources.
For that criterion and sample, 240 (83\%) of Class 0$+$I YSOs have detections,
while 112 (60\%) of Flat SED sources have detections, and 28 (40\%) of
Class II sources have detections. 
To estimate the number of Stage 0$+$I sources based on detection of the line,
we could add the Flat SED  and Class II sources that are detected to the 
list of Stage 0$+$I sources, yielding a number of 380 potential Stage 0$+$I sources
out of 546 observed sources. 

If we assume that our sample of observed sources was not biased,
we can use the detection percentages to extrapolate to the number that would be
detected in the full catalog. This analysis yields numbers of
plausible detected sources of 271 Class 0$+$I sources and 125 Flat SED
sources for a total of 396 likely detected sources within those
two classes (Table~\ref{summarytab}).  We do not include the Class II sources in the
extrapolated statistics because the Class II sample
was clearly biased.  

In contrast, the selection of Class 0$+$I and Flat SED sources was not biased
by the values of $\alpha$, strength of infrared emission, or any other variable
known to us. The quite small fraction of sources that remained unobserved 
resulted entirely from time limitations.

These numbers are certainly an overestimate for the
following reasons.
\begin{enumerate}
\item Some sources
without their own dense envelopes may be superimposed on emission
from dense envelopes of other sources. This possibility can be examined
only with mapping and detailed analysis, which was not possible for this
large sample. The study by Carney et al. (in prep.) with mapping of
a smaller sample will allow some estimates of the size of this
effect.
\item Weak lines may be associated with very low mass remnant envelopes
or even disks. While a quantitative relation between \hcop\ emission and
mass has not been established, we will follow the example of 
\citet{2009A&A...498..167V}, who set a minimum value for the integrated 
intensity (\S \ref{threshold}). This threshold was intended to eliminate Stage II sources (disk-only), which may have \hcop\ emission.
\end{enumerate}

\subsection{Results for a Threshold Emission Strength}\label{threshold}

%van Kempen et al.
\citet{2009A&A...498..167V} applied a series of requirements to decide
if sources were in Stage 0$+$I. The main requirement was that
the \hcop\ was extended, but peaked on the source with an integrated intensity,
 \imb\ $> 0.4$ \kkms. If the \hcop\ was not extended, an alternative
criterion required continuum emission from dust at 850 \micron\ to be
extended, but peaked on the source. Because we do not have maps,
the only criterion we can easily apply is one on the strength of the
\hcop\ \jj32\ emission. The criterion used by \citet{2009A&A...498..167V}
was that  \imb\ $> 0.4$ \kkms, for the \jj43\ line.
To translate this limit to our slightly more easily
excited line, we compared values of   \imb\ for our \jj32\
lines with a sample of 27 sources also observed in the \jj43\ line by 
Carney et al. (in prep.).  The \jj43\ maps were convolved to the 28\arcsec\ beam size of the CSO single pointing observations and the spectra were fitted
 using a single line Gaussian profile.
For that sample our values of   \imb\ are $1.70\pm 0.10$ times higher,
where the uncertainty is the standard deviation of the mean. Therefore,
the equivalent   \imb\ for the \jj32\ line to the
\citet{2009A&A...498..167V} criterion is  \imb\ $\ge 0.68$ \kkms.  We have also compared our sample to the Ophiuchus sample observed by \citet{2009A&A...498..167V} and found 17 sources with detections that overlap in both samples.  If we include these sources in our average line ratio, we find that our \jj32\ line measurements are 1.58$\pm$0.10 times higher than the \jj43\ integrated intensities.  We note that these maps are at a higher resolution and are not convolved to the same beam size of our measurements.  With this caveat in mind, we will use the \imb\ $\ge 0.68$ \kkms\ criteria for the threshold in our analysis, but note in the following paragraph the differences. 
As noted above, the threshold was intended to discriminate against \hcop\ emission from disks; disk emission would be highly diluted in our beam, so our scaled up threshold is a conservative choice that probably excludes some
Stage 0$+$I sources with low-mass envelopes.

What changes if we apply a cut to the integrated intensity so that only
sources with a value above that threshold would qualify as Stage 0$+$I sources?
With this criterion, there are 208/288  (72\%)  and 90/188 (48\%)  of the observed Class 0$+$I sources and
Flat SED sources with   \imb\  above that threshold, respectively. The status of Flat SED sources has never been
very clear, but they have been assumed to be a transitional class.
The fact that roughly half achieve Stage 0$+$I status by the standards
we are using is consistent with that picture.
Interestingly, 21 Class II sources (30\% of the observed sources) also
satisfy that criterion.  If we instead used the slightly lower \imb\ threshold average of 0.63 \kkms\ from both the Carney et al. (in prep.)  and \citet{2009A&A...498..167V} samples, we find 13 additional sources above the threshold  or  213/288  (74\%) of Class 0$+$I,   95/188 (51\%) Flat SED, and 24 Class II sources. The detected Class II 
sources have values of $\alpha$ that are highly skewed toward
the boundary between Class II and Flat SEDs. Since these were originally
classified as either Class 0$+$I or Flat (to be included in our sample), they
could be Stage 0$+$I sources with envelopes that are less opaque or
that are viewed closer to face-on. They could also be caused primarily
by contamination (projected on \hcop\ emission from another source or
extended emission); comparison to detections of submillimeter continuum
emission (see \S \ref{smm}) suggest the latter explanation.

The number of sources detected above threshold is listed for each cloud in column 12 of Table \ref{tbl-2}. 
As in \S \ref{any}, we extrapolate to the full sample for Class 0$+$I
and Flat SED sources, but not for Class II sources.  These results are summarized
in Table \ref{summarytab}.

\subsection{Comparison to Far-Infrared or Submillimeter Continuum 
Emission Detections}\label{smm}

Detection of submillimeter continuum emission concentrated on an infrared
source, but extended more than would be expected of a disk, is another
criterion often used to identify Stage 0$+$I sources 
(e.g. \citealt{2009A&A...498..167V}). We do not have a complete survey
of our sample for submillimeter continuum emission, but we can compare
to the 167 sources with submillimeter observations within our sample, as tabulated by 
%Dunham et al.
\citet{2013AJ....145...94D}.
Table \ref{smmcomp} gives the numbers for various combinations of detections
above threshold of \hcop\ and a far-infrared or submillimeter continuum 
emission detection in a wavelength range $70 \micron < \lambda < 850 \micron$.

Of the sources in our sample with submillimeter detections, 
91 of those are detected 
in \hcop\ and 4 are undetected, 85 above the threshold of $0.68$ \kkms.
Only 4 sources with submillimeter emission have no \hcop\ detection and
6 have a weak detection (\imb\ $ < 0.68$ \kkms ). There is thus generally strong
agreement between the \hcop\ and submillimeter criteria. There are
however, 55 sources with strong \hcop\ detections and submillimeter data,
but no submillimeter detection. Without more information on the sensitivity
of the submillimeter data, it is hard to decide if there is real disagreement.

Very few sources with submillimeter continuum detections are undetected in
\hcop. Since the sample with submillimeter detections is different, there
are 35 sources with submillimeter detections that we did not observe.
If one added all the sources with submillimeter detections to the list of
Stage 0$+$I candidates, the numbers and timescales would increase, but not
substantially.

We can also examine the changes in the agreement between \hcop\ and 
submillimeter continuum emission with respect to SED class.
There are 52 Class 0$+$I and Flat SED sources that meet the Stage 0$+$I 
emission criteria but do not have submillimeter emission and 162 
sources that have not been observed in the submillimeter.
There is only 1 submillimeter detection  out of the 21 Class II sources that
meet the \hcop\ Stage 0$+$I criteria.  That fact clearly suggests that 
\hcop\ detections toward Class II sources are likely to be due to
contamination. We include all the numbers so that readers can make
their own judgments, but we favor the values in Table \ref{summarytab}
that are extrapolated to the full sample but without Class II sources.

\subsection{Timescales for Stage 0$+$I}\label{timescales}
The standard calculation for how long is spent in a given class 
uses the number of sources in that class, the number of sources in a 
reference class, and an assumed timescale for that reference class
(e.g., \citealt{2009ApJS..181..321E}).
The reference class has generally been taken to be Class II
objects. While some Class II sources do have \hcop\ emission, it may
be caused by confusion. In the absence of maps that could test that possibility,
we will continue to
associate Class II sources with Stage II sources, those without substantial
envelopes but with infrared excess indicative of a disk. Therefore
we will calculate the duration of the Stage 0$+$I phase using the following equation.

\begin{equation}
\Delta t({\rm Stage\ 0+I}) = \Delta t({\rm Class\ II}) \frac{N({\rm Stage\ 0+I})}
{N({\rm Class\ II})}
\end{equation}

Table \ref{summarytab} shows the numbers of sources and associated
timescales for various possible calculations. The timescales in this
table assume a Class II duration of 2 Myr for consistency with
previous work (but see below). We show results for any detection and
for detections above the threshold,   \imb\ $ \geq 0.68$ \kkms.
The numbers indicated by ``Raw" count
only actual detections but include any class. The numbers indicated
by ``Extrapolated" account for the incomplete sampling of the full
catalog, using the fraction observed, but only in Class 0$+$I and Flat SED,
because the Class II sample was highly incomplete and very biased.

With 1243 Class II YSOs in the clouds, the estimates in Table \ref{summarytab}
for $\Delta t({\rm Stage\ 0+I})$ range from 0.51 to 0.62 Myr.
As discussed above, we recommend the value based on extrapolation to the
full sample, but excluding Class II sources; in this case the 
timescale for Stage 0$+$I is 0.54 Myr for a Class II timescale of 2 Myr.
This timescale is very similar to that found for Class 0$+$I sources alone
by \citet{2009ApJS..181..321E} for the c2d clouds and slightly larger
than timescales found for Class 0$+$I sources in the Taurus and Orion clouds
when similar criteria were used 
% Dunham PPVI
\citep{2014prpl.conf..195D}.
The loss of some Class 0$+$I objects has been compensated by the addition
of some Flat SED objects, but now we have a better estimate for
the timescale of the entire embedded phase.

As noted above, in previous analyses,
we have assumed that the fraction of YSOs with infrared excess declines
exponentially with a half-life of $2 \pm 1$ Myr
% Mamajek
\citep{2009AIPC.1158....3M}.
However, these estimates depend entirely on
the timescales for loss of infrared excess.
Recently, longer timescales have been suggested by studies
of older associations 
%(PPVI chapter on stellar ages),
\citep{2014prpl.conf..219S},
and models of disk evolution suggest a relatively constant and
high fraction of disks up to about 2.5 Myr, 
followed by a rapid decrease from 2.5 to 4 Myr 
%(Alexander and Armitage 2009).
\citep{2009ApJ...704..989A}.
Observations accounting for more rapid dissipation of disks around
close binaries are broadly consistent with this model (Kraus et al. 2012).
Consequently, the timecales for Stage 0$+$I could easily be 1.5-2 times 
longer. 

Conversely, the timescales could be shorter if we added in
the denominator in equation 2 the Class III sources with 
infrared excesses as seen by {\it Spitzer} surveys. 
If these are also counted in the statistics of infrared excess in clusters,
then the half-life from cluster studies includes that fraction of Class III
objects. Early studies of disk fraction in clusters were less sensitive
to small excesses than was {\it Spitzer}, but recent studies by 
%Ribas 14
\citet{2014A&A...561A..54R},
using Spitzer data on 22 associations, provide the most comparable 
sample. They find that the timescale for infrared excess depends
on the wavelength where the excess first appears and
on stellar mass
%Ribas 15
\citep{2015arXiv150200631R}.
The timescales in the last reference, 
converted from $1/e$ time to half-life, range from 1.9
to 3.0 Myr, with the larger time corresponding to excesses beginning at
wavelengths as long as 20 \micron. 
The latter is probably the best value for comparison
if we include our Class III sources.
There are 1188 Class III objects in the tables of YSOs from 
\citet{2013AJ....145...94D}. Some infrared excess at wavelengths out
to 24 \micron\ was required for these
to become YSO candidates. If we include those in equation 2 and take a
3.0 Myr half-life, the timescale for Stage 0$+$I drops to 0.41 Myr.  We note, however, that our class III sample may be $\sim$50--100\%  contaminated 
by background asymptotic giant branch (AGB) stars (Dunham et al., in prep.).  For example, if the number of class III sources had at least 50\% AGB star contamination, we would still obtain a similar Stage 0$+$I timescale of 0.55 Myr.

\subsection{Correlations}\label{corr}

In this section, we consider how   \imb\ correlates with various quantities
used to trace evolution. If $\alpha$ decreases and \tbol\ increases as the
envelope mass decreases, one might expect correlations (positive and negative
respectively) of the \hcop\ emission with these quantities.
However, the emission from \hcop\ is not simply proportional to the mass of
the envelope. Warmer and denser gas will emit more strongly. 
We have no direct measure of those quantities, 
but the gas temperature will be higher around sources of higher luminosity.

\begin{figure}
\center
\includegraphics[scale=0.35, angle=0]{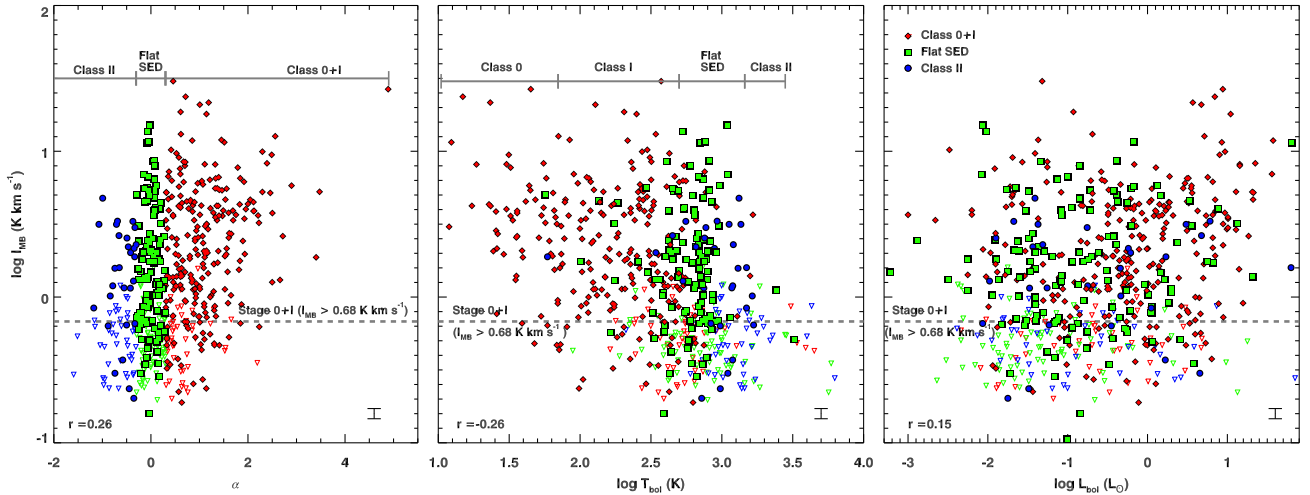}
\caption{
The integrated intensity of lines is shown versus $\alpha$, \tbol, and \lbol, separated by Class as determined from $\alpha$ (Class 0$+$I, Flat SED, and Class II are red diamonds, green squares, and blue circle solid points, respectively). The left and middle panel show differences between using $\alpha$ and \tbol\ to classify sources (grey bars). Binned averages and their respective relations are shown by the yellow stars and errors. Pearson correlation coefficients and mean errors in \imb\ are also shown (see \S \ref{corr}). \hcop\ \jj32 upper limits are shown as downward triangles. 
}
\label{imb}
\end{figure}

Figure~\ref{imb} plots  \imb\ versus the spectral index, $\alpha$, \tbol, 
and \lbol,
with inverted triangles indicating upper limits. 
In cases where lines had two velocity components, we
used the   \imb\ of the stronger line.
The grey solid lines show the $\alpha$ values that separate the SED classes.  

In the \imb\ versus \tbol\ plot, we can separate Class 0 from Class I,
and we also show the 
divisions between classes for extinction-corrected \tbol\ suggested by 
%Evans et al. 2009
\citet{2009ApJS..181..321E}:
\begin{description}
\item[0] $\tbol < 70$ K;
\item[I] 70 K $\leq \tbol\ \leq 500$ K;
\item[Flat] 500 K $< \tbol\ \leq 1450$ K;
\item[II] 1450 K $< \tbol\ \leq 2800$ K.
\end{description}

\begin{figure}
\center
\includegraphics[scale=0.36, angle=0]{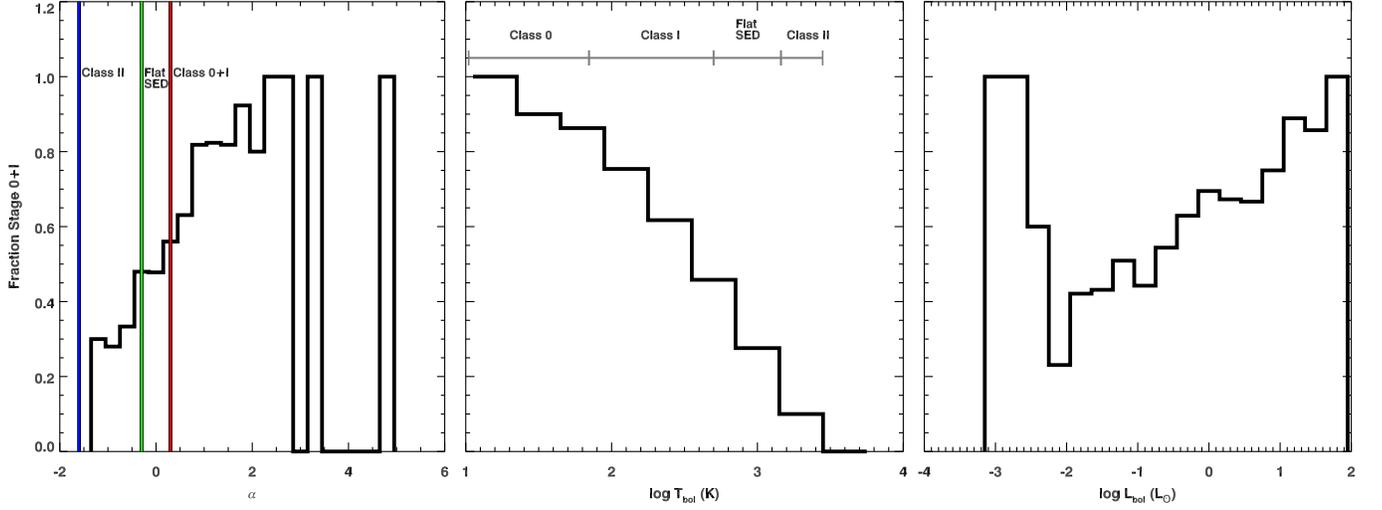}
\caption{
The fraction of Stage 0$+$I sources versus 
%$\alpha$, 
$\alpha$, \tbol, and \lbol. The vertical colored lines in the left panel mark the boundaries between 
Class 0$+$I, Flat SED, and Class II protostars by $\alpha$ and the middle panel show these classification boundaries as defined by \tbol\ (grey bars). The bins plotted at zero fraction above $\alpha = 2$ have no sources, so 
the fraction is undefined. Every source with $\alpha > 2$ is detected. The peak at low values of \lbol\
is misleading as these bins suffer from very low number statistics.} 
\label{det_hist}
\end{figure}

The data are very scattered, but there are trends for stronger emission 
with larger $\alpha$ (left panel), smaller \tbol, and possibly larger
\lbol. To test for a significant ($3 \sigma$) correlation, 
we require the absolute values of 
the Pearson correlation coefficient to exceed $3/\sqrt{N-1}$, where $N$ is the
number of data points. Including all detections, we have 377 data points
(3 have no values for $\alpha$), so we test for $|r| > 0.16$. The tests are done for $\log \imb$ versus $\alpha$, $\log \tbol$, and
$\log \lbol$. 
We find significant correlation for $\alpha$ ($r = 0.26$),
and an anti-correlation for \tbol\ ($r = -0.26$), 
but not for \lbol\ ($r = 0.15$). The  {\it fraction} of sources with detections above the threshold
shows a strong and smooth correlation
with evolutionary indicators.
Figure~\ref{det_hist}  shows the detection fraction of all sources above the
\imb\ threshold for Stage 0$+$I sources
in bins of $\alpha$, log \tbol, and log \lbol. 
The detection fraction
increases strongly with $\alpha$ through the flat category and is 80\% or
greater for sources with $\alpha > 0.75$.  The fraction with detections steadily increases as
\tbol\ decreases or \lbol\ increases.

\subsection{The Distribution of Stage 0$+$I Sources}\label{dist}

\begin{figure}
\center
\includegraphics[scale=0.5, angle=0]{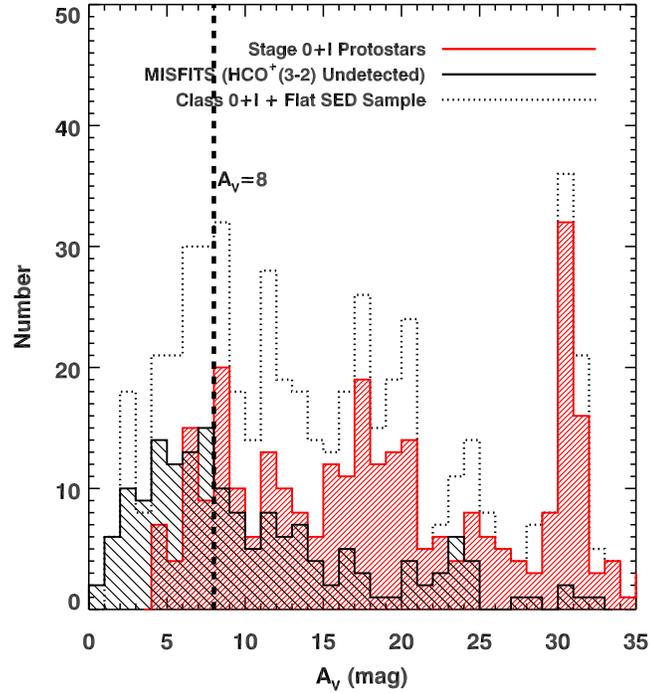}
\caption{
The number of sources that are SED Class 0$+$I or Flat (dotted line) compared to the distribution of Stage 0$+$I sources (red hashed histogram) and MISFITS (black hashed histogram).  We find 89\% of Stage 0$+$I protostars to lie above \av$=8$ mag (black vertical dashed line) compared to 75\% of Class 0$+$I or Flat SED YSOs in the catalog. MISFITS are evenly distributed on either side of \av$=8$ mag. } 
\label{stage1hist}
\end{figure}

\begin{figure}
\center
\includegraphics[scale=0.7, angle=0]{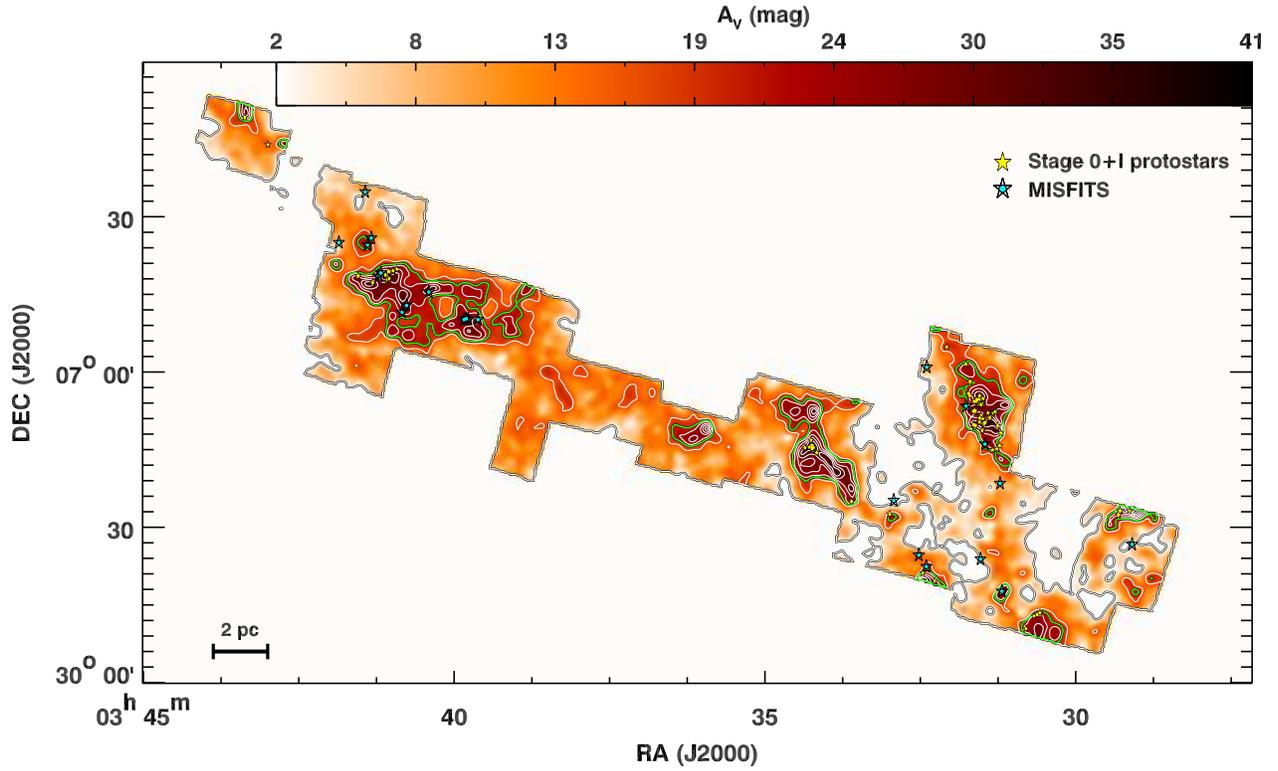}
\caption{The distribution of  bona fide Stage 0$+$I protostars and MISFITS in the Perseus molecular cloud. The background image is the 180$\arcsec$ resolution extinction map with contours ranging from 2 to 41 mag in intervals of 4 mag. The black contour outlines the extinction map at the completeness limit of $\av = 2$ mag and the green contour shows where $\av = 8$ mag. The yellow stars are Stage 0$+$I protostars. Sources that are filled cyan circles and open stars correspond to MISFITS that are undetected in \hcop\ \jj32\ emission. 
}
\label{cloud}
\end{figure}

Based on preliminary results of this survey for \hcop,
\citet{2010ApJ...723.1019H} showed that Class 0$+$I and Flat SED sources
were strongly concentrated into regions with large-scale extinction
above 7 or 8 mag, and that this statement became stronger when
sources without \hcop\ emission (MISFITS) were removed.
We can now revisit this result with the full sample.
Figure~\ref{stage1hist} show the \av\ at the source position, determined
from background star extinctions averaged over 270\arcsec, for both Stage 0$+$I 
and all 535 Class 0$+$I and Flat SED YSOs in the sample. 
The sample of all Class 0$+$I and Flat SED sources (dotted histogram) are concentrated toward regions
of high extinction, with 75\% of all Class 0$+$I and Flat SED sources found
toward $\av > 8$ mag (black vertical dashed line).  
Stage 0$+$I sources (black hashed histogram) are still more strongly concentrated in regions of 
high extinction with 89\% of them above $\av = 8$ mag and the MISFITS (red hashed histogram) are distributed evenly on either side of $\av = 8$ mag line.
Figure~\ref{cloud} shows the distribution of bona fide Stage 0$+$I protostars and MISFITS in the Perseus molecular cloud. The background image is the extinction map with a 180\arcsec\ resolution and contours ranging from 2 to 41 in intervals of 4 mag. The black contour outlines the extinction map at the completeness limit of $\av = 2$ mag and the green contour shows where $\av = 8$ mag. The yellow stars are Stage 0$+$I protostars. Sources that are filled cyan circles with an open star correspond to MISFITS that are undetected in \hcop\ \jj32\ emission. 

It is important to recall that the extinction maps are based on star counting
and have much poorer spatial resolution than the \hcop\ observations.
High extinction is not just a reflection of the local core density.
In fact, one might expect some MISFITS at high extinction, which would
distort the mid-infrared SED, despite our efforts to correct for it.
Indeed there are a few such objects in the histogram, but the main
trend is for a higher \hcop\ detection fraction in regions of higher
large-scale extinction.

\subsection{Properties of the MISFITS}\label{misprop}

\begin{figure}
\center
\includegraphics[scale=0.4, angle=0]{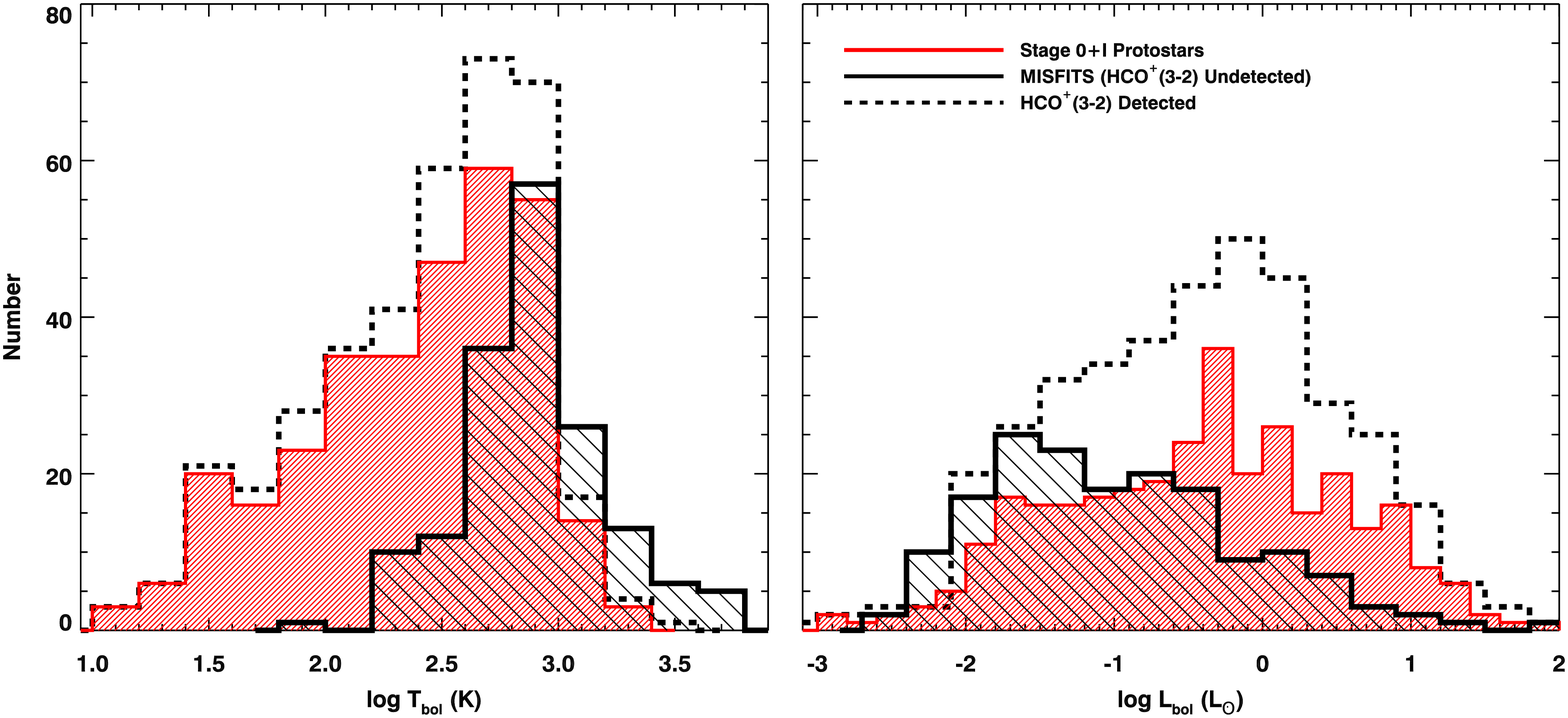}
\caption{The number of MISFITS (undetected in \hcop\ \jj32; hashed histogram) are compared to Class 0+1 and Flat SED YSOs that are detected in \hcop\ \jj32 versus 
\tbol, and \lbol. Distributions are skewed to the right. The median values for \tbol\ and \lbol\ are 353 K and 0.43 K for Stage 0+I, 774 K  and 0.083 \lsun\  for MISFITS, and 330 K and  0.41 \lsun\  for all detected YSO distribution, respectively. }
\label{properties}
\end{figure}

We can explore trends in properties of the MISFITS or YSOs undetected in \hcop\ \jj32\,. In Figure~\ref{properties}, we compare the number of MISFITS (black hashed histogram) to Class 0+1 and Flat SED YSOs that are detected in \hcop, and Stage 0$+$I protostars with strong \hcop\ emission versus \tbol\ and \lbol. In all cases, the distributions for Stage 0$+$I sources are skewed with respect to those of the MISFITS.     The median values for \tbol\ and \lbol\ are 353 K and 0.43 \lsun\ for Stage 0$+$I sources (detected above threshold), 774 K  and 0.083 \lsun\  for MISFITS, and 330 K and  0.41 \lsun\  for  YSOs detected at any level, respectively. In general, Class 0$+$I and Flat SED YSOs that are undetected in \hcop\ have a factor of $\sim2$ higher \tbol\ and a factor of $\sim$5 lower \lbol\ than detected YSOs. 

\begin{figure}
\center
\includegraphics[scale=0.5, angle=0]{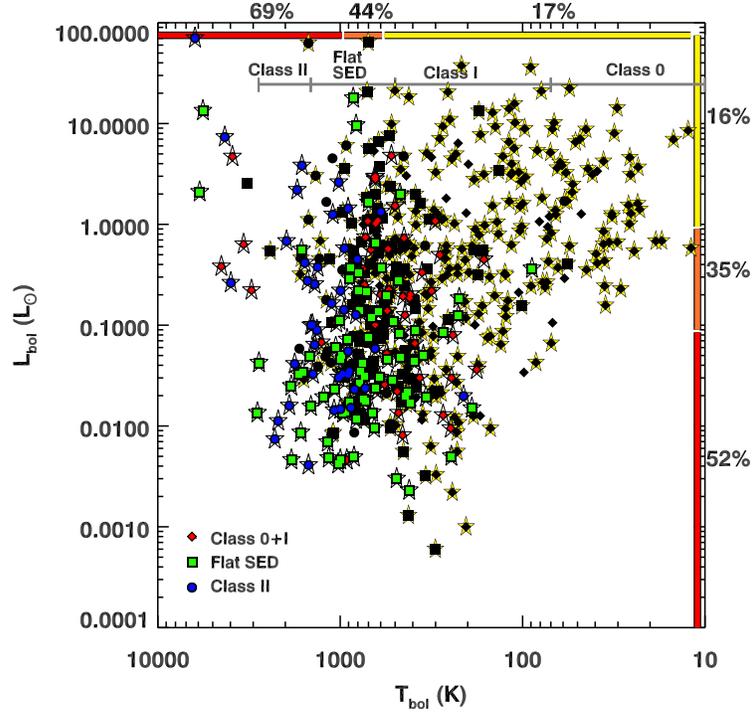}
\caption{Bolometric luminosity (\lbol) is shown versus the bolometric temperature (\tbol) for all YSOs observed in  \hcop\ \jj32\, separated by Class as determined from $\alpha$ (Class 0$+$I, Flat SED, and Class II are  diamonds,  squares, and  circle solid points, respectively). Undetected YSOs or MISFITS are shown in color (Class 0$+$I, Flat SED, and Class II are  red,  green, and  blue solid points, respectively) with an open star.  Stage 0+I protostars are indicated by the yellow stars. The percent of YSO contamination defined as the number of MISFITS divided by the total surveyed in a bin is shown color coded from more (red) to less (yellow). Less contamination ($\leq 17$\%) is found where \tbol$\lesssim$ 600 K and \lbol$\gtrsim$ 1 \lsun.}
\label{misproperties}
\end{figure}

\begin{figure}
\center
\includegraphics[scale=0.5, angle=0]{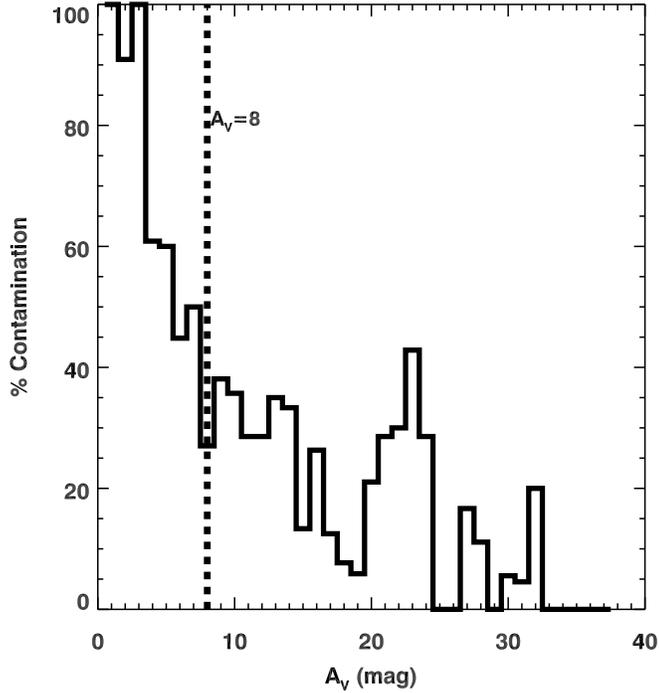}
\caption{The percent of YSO contamination defined as the number of MISFITS (undetected in \hcop\ \jj32) divided by the total number surveyed versus $\av$.  The contamination fraction is $>90$\% at \av$<$4 mag and decreases as \av\ increases. The rise above $\av=20$ mag is due to low number statistics. }
\label{contamination}
\end{figure}

Figure~\ref{misproperties} shows \lbol\ versus \tbol\ for all YSOs observed in  \hcop\  separated by Class as determined from $\alpha$ (Class 0$+$I, Flat SED, and Class II are  diamonds,  squares, and  circle solid points, respectively). Undetected YSOs or MISFITS are shown in color (Class 0$+$I, Flat SED, and Class III are  red,  green, and  blue solid points, respectively) with an open star.  Stage 0$+$I protostars are indicated by the yellow stars. The percent of YSO contamination defined as the number of MISFITS divided by the total surveyed in a bin is shown color coded from more (red) to less (yellow). Low ($< 17$\%) contamination is found where \tbol$\lesssim$ 600 K and \lbol$\gtrsim$ 1 \lsun. We further explore the percent of YSO contamination versus $\av$ in Figure~\ref{contamination}.  The contamination fraction is $>90$\% at $\av < 4$ mag and decreases with increasing \av. The rise above $\av=20$ mag is due to low number statistics.

% Table 3  8/25/14 ALH
% 082514 renamed table 4 NJE
% 01/18/15 ALH updated 
\begin{deluxetable}{lc}
%\tabletypesize{\scriptsize}
\tablecaption{Stage 0+I Sources vs. Submm detections\label{smmcomp}}
\tablewidth{0pt}
\tablehead{
\colhead{Criteria} &
\colhead{Number}\\ 
\colhead{(1)} &
\colhead{(2)} \\
%\colhead{(4)} \\
 }
\startdata
Strong line detection\tablenotemark{a}&  319 \\
Strong line detection $+$ Submm detection & 85\\
Strong line detection or Submm detection & 329\\
Strong line detection $+$ Submm undetected & 55\\
Weak line detection $+$ Submm detection & 6\\
\hcop Undetected $+$ Submm detection & 4\\
Source not in \hcop survey $+$ Submm detection & 35\\
\enddata
\tablenotetext{a}{Strong line defined as a source meets the $I_{\rm MB}>0.68$ \kkms\ detection critiera. 
}
%\end{deluxetable*}
\end{deluxetable}

% Table 3  08/25/14 NJE
% 102214 NJE edit
%011815 ALH corrections
%111914 NJE edit to fit into one column

\begin{deluxetable}{llccl}
%\tabletypesize{\scriptsize}
\tablecaption{Numbers and Timescales for Stage 0+I Sources 
\label{summarytab}}
\tablewidth{0pt}
\tablehead{
\colhead{Quantity} &
\colhead{Sample} & 
\colhead{Detected} & 
\colhead{\imb\ $> 0.68$} &
\colhead{Include}  \\
\colhead{} &
\colhead{} &
\colhead{} &
\colhead{$\kkms$} &
\colhead{Class II?} \\
\colhead{1} &
\colhead{2} &
\colhead{3} &
\colhead{4} &
\colhead{5} \\
 }
\startdata
Number & Raw  & 380  & 319 & Yes \\
Number & Extrapolated  & 396  &  335 & No \\
Timescale & Raw  & 0.62 Myr  & 0.51 Myr & Yes \\
Timescale & Extrapolated & 0.64 Myr  & 0.54 Myr & No \\
\enddata
%\tablenotetext{a}{ TBD }
%\end{deluxetable*}
\end{deluxetable}

% summary.tex
% 102314 NJE initiated
% 111014 ALH edits
% 012015 ALH
% 012915 NJE
% 020515 NJE

\section{summary}

The goal of this work was to identify Stage 0$+$I sources (those 
with a substantial envelope of dense gas) among the larger
sample of Class 0$+$I and Flat SED sources. To achieve this,
we have surveyed for emission in the \jj32\ line of 
\hcop\ toward 546 sources in nearby clouds. The sample is
based on the YSO lists for the combined c2d and Gould Belt
surveys with Spitzer. It is focused on Class 0$+$I and Flat SED sources,
but a few Class II sources were included. The surveyed sources
represent 88\% of the Class 0$+$I sources and 90\% of the Flat SED
sources in the full list of YSOs in \citet{2013AJ....145...94D}.

If we apply a threshold integrated intensity of
$\imb \geq 0.68$ \kkms\ to define a Stage 0$+$I source,
72\% of the Class 0$+$I and 48\% of the Flat SED sources would
be classified as Stage 0$+$I sources (\S~\ref{threshold}). A few Class II sources were also
detected, but these sources were originally classified as Flat or
Class 0$+$I sources, so are very unrepresentative of the much larger
group of Class II sources. If we exclude them, but extrapolate
(modestly) to the full c2d plus Gould Belt sample, we predict
335 Stage 0$+$I sources in all the clouds.  

Based on a 2 Myr timescale for Class II sources, the most 
appropriate timescale for Stage 0$+$I (the entire embedded phase)
would be 0.54 Myr (\S~\ref{timescales} and Table~\ref{summarytab}). Other choices for how to calculate this
number are certainly possible, and we have tried to supply enough
information so that other calculations could be made.

There is a good, but imperfect, agreement between \hcop\ detections
and far-infrared or submillimeter continuum emission (\S~\ref{smm}), within the limited sample with
both kinds of data available .

The \hcop\ detection fraction increases with spectral index ($\alpha$)
and bolometric luminosity (\lbol), and decreases with bolometric
temperature (\tbol), all trends that one might expect (\S~\ref{corr}). 
In particular, nothing distinctive occurs within the Flat SED
category, and about half of those sources qualify as Stage 0$+$I sources,
suggesting that this category has no physical significance.
Unlike the rather smooth dependence of the detection fraction on
the usual evolutionary indicators,
the {\it strength} of the \hcop\ emission scatters widely
but trends versus these possible evolutionary indicators are clearly
significant.
Other variables clearly affect the \hcop\ emission strength, but
the correlation of emission strength with \lbol\ is not signficant.

The concentration of Class 0$+$I and Flat SED sources to regions of
high extinction seen in our previous study \citep{2010ApJ...723.1019H}
is strengthened when consideration is narrowed to Stage 0$+$I sources (\S~\ref{dist}).

YSOs undetected in \hcop\ lie preferentially at higher \tbol\ and at 
lower \lbol\ than detected YSOs.
The YSO contamination, defined as the number of undetected YSOs divided by the total surveyed, is low where \tbol$\lesssim$600 K and \lbol$\gtrsim$ 1 \lsun. The contamination fraction is $>90$\% at $\av < 4$ mag and decreases with increasing \av.

\medskip
% ack.tex
% 08/12/14
% 111914

We thank M. Dunham, J-H. Chen, H-J. Kim, J. Green, C. Salyk,
M. Merello, N. Vutisalchavakul, I. Oliveira, M. Song, and E. Yu
for assistance in obtaining observations at the CSO 
and A. Gusdorf, G. Siringo, and J. Forbrich for assistance at APEX. 
We also thank M. Carney and E. van Dishoeck for providing their \hcop\
\jj43\ data for comparison in advance of publication and T. Huard for informative discussions.
ALH is supported by an NSF AAPF under award AST-1302978.
NJE is supported by NSF Grant AST-1109116 to the University of Texas at Austin.
NJE thanks the European Southern Observatory, Santiago,
for hospitality during an extended visit when some of this work was done. This work was partially carried out in the Max-Planck Research Group {\it{Star formation throughout the Milky Way Galaxy}} at the Max-Planck-Institut f\"{u}r Astronomie.

\clearpage

\bibliographystyle{apj}

\bibliography{njebib,more,alh}
\clearpage
% Table 2
% 090314 NJE updated after redoing average spectra
%111314 ALH updated with corrections and table comments
\begin{deluxetable}{lrrrrrrrrrrrl}
\tabletypesize{\scriptsize}
\tablecaption{\hcop\ \jj32\ Detection Statistics for Clouds\tablenotemark{a}
\label{tbl-2}}
\tablewidth{0pt}
\tablehead{
 \colhead{Cloud} & \colhead{D} & \colhead{Size} & 
 \colhead{0$+$I} & \colhead{0$+$I} & 
 \colhead{Flat} & \colhead{Flat} &
 \colhead{II} & \colhead{II} &
 \colhead{V} & \colhead{$\Delta$V} &
 \colhead{Number} & \colhead{Notes}  \\
 \colhead{}  & \colhead{(pc)} & \colhead{(pc)}  &  
 \colhead{Y} & \colhead{N}  & 
 \colhead{Y} & \colhead{N} & 
 \colhead{Y} & \colhead{N} & 
 \colhead{(\kms)} & \colhead{(\kms)} &  
 \colhead{Stg 0$+$1\tablenotemark{b}}  & \colhead{} \\
  \colhead{(1)}& 
    \colhead{(2)}& 
    \colhead{(3)}& 
    \colhead{(4)}& 
    \colhead{(5)}& 
    \colhead{(6)}& 
    \colhead{(7)}& 
        \colhead{(8)}& 
    \colhead{(9)}& 
    \colhead{(10)}& 
    \colhead{(11)}& 
    \colhead{(12)}& 
    \colhead{(13)}\\
  }
\startdata
Aquila       & 260 & 7.56 &           66 &           15 &           39 &           21 &           13 &           26 & 7.4  & 3.0 &          97 & \\
Auriga/CMC       & 450 & 5.98 &           28&            5 &            5 &            3 &            6 &            1 & €"0.9 & 1.8 &           30 & \\
Auriga North & 450 & 1.31 &            1 &            0 &            0 &            0 &            0 &            0 & -6.9 & 1.9 &            1 & \\
Cepheus 1    & 300 & 1.75 &            5 & 0            &            2 &            1 &            0 &            1 & -4.4 & 1.3 &            5 & stronger \\
  &   &   &   &   &   &   &   &   & -2.9 & 1.0 &   & weaker \\
Cepheus 3    & 288 & 1.75 &            4 &            2 &            0 &            1 & 0            &            1 & 2.6 & 1.3 &            4 & \\
Cepheus 4    & 325 & 1.08 &            0 &            0 &            0 &            1 & 0            &            0 & \ldots & \ldots &        0 & \\
Cepheus 5    & 200 & 1.07 &            0 & 0            &            3 &            0 &            0 &            0 & -8.0 & 1.4 &            3 & \\
Chamaeleon I & 150 & 1.30 &            3 &            1 &            1 &            4 &            2 &            5 & 4.8  & 1.7 &            2 & \\
Chamaeleon II & 178 & 1.78 &            1 &            1 &            0 &            0 &            1 &            0 & 2.9  & 0.8 &            1 & \\
Corona Australis (CrA) & 130 & 0.98 &        5 &            1 &            2 &            2 &            0 &            1 & 5.2  & 2.9 &            7 & \\
IC5146 E     & 950 & 4.42 &            8 &            5 &            0 &            2 &            0 &            3 & 6.5  & 1.2 &            7& \\
             &     &      &              &              &              &              &              &              & 8.1  & 1.3 &              & stronger \\
IC5146 NW    & 950 & 5.28 &           13 &            0 &            3 &            1 &            0 &            0 & 4.2  & 2.0 &           11 & \\
Lupus I      & 150 & 1.68 &            0&            1 &            0 &            2&            0 &            0 & 4.9  & 0.8 &            0 & \\
Lupus III    & 200 & 2.22 &            1 &            1 &            0 &            5 &            0 &           1 & \ldots & \ldots &        1 & \\
Lupus IV     & 150 & 0.90 &            0 &            1 &            0 &            0 &            0 &            0 & \ldots & \ldots &       0 & \\
Lupus VI     & 150 & 1.46 &            0 &            1 &            0 &            0 &            0 &           0 & \ldots & \ldots &        0 & \\
Musca        & 160 & 1.47 &            0 &            1 &            0 &            0 &            0 &           0 & \ldots & \ldots &        0 & \\
Ophiuchus    & 125 & 3.08 &           24 &            3 &           26 &           14 &            1 &            2 & 3.1  & 1.8 &           45 & stronger \\
         &     &      &              &              &              &              &              &              & 4.6  & 1.3 &              & \\
Perseus      & 250 & 4.83 &           66 &            7 &           18 &           16 &            1 &            0 & 5.3 & 2.0 &           77& \\
  & 250 &   &   &   &   &   &   &   & 7.8 & 2.7 &   & stronger \\
Scorpius/OphN & 130  & \ldots &            0 &            1 &            0 &            1 & 0 & 0 & \ldots & \ldots & 0 & \\
Serpens      & 429 & 3.92 &           15 &            2 &           13 &            2 &            4 &            1 & 7.6 & 2.1 &           28 & stronger \\
  &   &   &   &   &   &   &   &   & 9.6 & 1.6 &   & \\
\hline\hline
Total\tablenotemark{c}	     &     &	   &       240  &       48    &           112 &           76&           28 &           42 &     &     &           319&  \\
Fraction\tablenotemark{d}  &     & 	   &      0.83	 & 	0.17       &          0.60 & 	0.40      &   0.40     & 0.60	            & 	  &     &          0.58 &
\enddata
\tablenotetext{a}{ Only clouds with at least 1 observed source are included in the table. }
\tablenotetext{b}{ Counts only detections above the threshold value of
integrated intensity.}
\tablenotetext{c}{ Total number of sources separated by class that are detected or undetected or are classified as Stage 0$+$1.}
\tablenotetext{d}{ Fraction of sources in each column compared to total
number surveyed.}
\tablecomments{ Columns are :
(1) Cloud Name;
(2) Cloud Distance in parsec;
(3) Average Cloud Size in parsec;
(4) Number of detected Class 0$+$I sources;
(5) Number of undetected Class 0$+$I sources;
(6) Number of detected Flat SED sources;
(7) Number of undetected Flat SED sources;
(8) Number of detected Class II sources;
(9) Number of undetected Class II sources;
(10) Cloud velocity (see Section~\ref{appendix});
(11) Cloud line width (see Section~\ref{appendix});
(12) Number of cloud sources classified as Stage 0$+$1;
(13) Comparative notes on average cloud \hcop spectra.
}
%\end{deluxetable*}
\end{deluxetable}

% Figures
% figs.tex
% 021714 NJE initialized
% 090314 NJE edited for new figs 1-3
%111314 ALH edits, fig updates
%011815 ALH edits, fig updates
% 012915 NJE

\clearpage

%Fig. 1
\begin{figure}
%\center
\includegraphics[scale=0.5, angle=0]{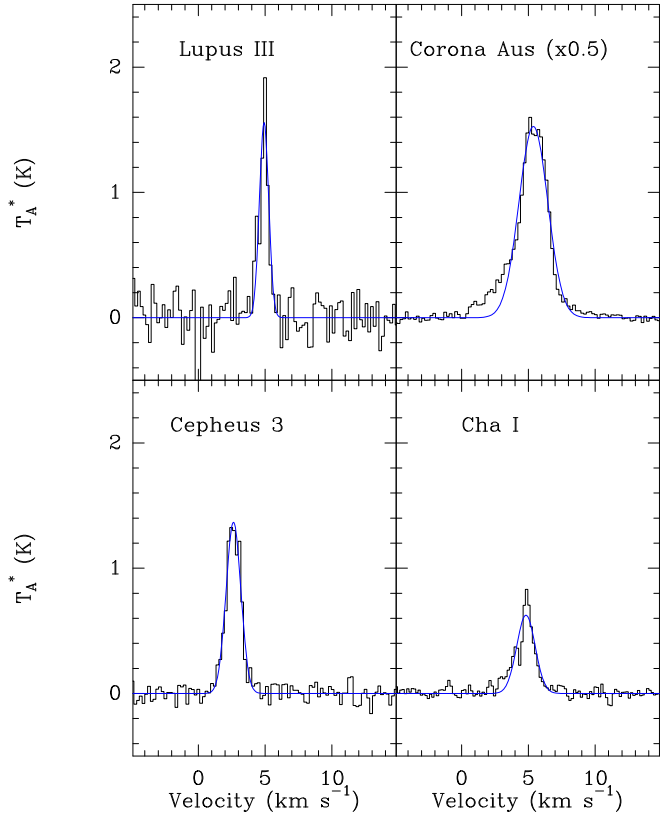}
\caption{
Average \hcop\ \jj32\ spectra of detected sources in four clouds.
The black histogram is the average spectrum.
The blue line is the fit to the spectrum (either one or two Gaussians).
The spectrum of Corona Aus has been scaled down by a factor of 2.
}
\label{spec1}
\end{figure}

%Fig. 2
\begin{figure}
%\center
\includegraphics[scale=0.5, angle=0]{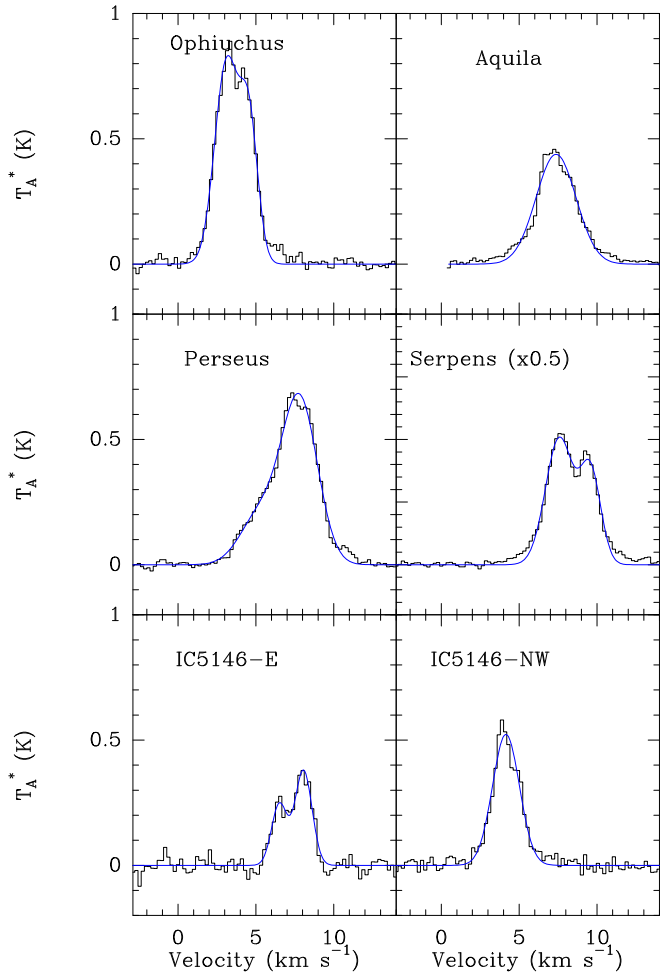}
\caption{
Average \hcop\ \jj32\ spectra of detected sources in six clouds.
The black histogram is the average spectrum.
The blue line is the fit to the spectrum (either one or two Gaussians).
}
\label{spec2}
\end{figure}

%Fig. 3
\begin{figure}
%\center
\includegraphics[scale=0.5, angle=0]{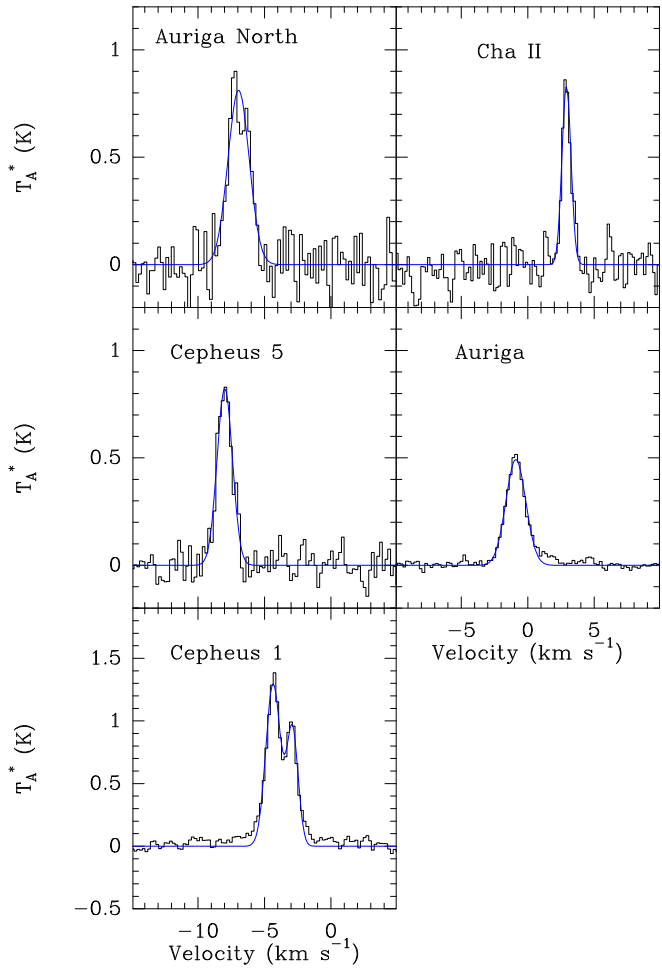}
\caption{
Average \hcop\ \jj32\ spectra of detected sources in five clouds.
The black histogram is the average spectrum.
The blue line is the fit to the spectrum (either one or two Gaussians).
The vertical scale for Cepheus 1 is larger than for other clouds.
}
\label{spec3}
\end{figure}

%Fig. 4

%Fig. 5

%Fig 5

%Fig. 6

%Fig. 7

%Fig 8

%Fig 9

%Fig 10

%Fig 11

\clearpage
%appendix.tex
% 090314 NJE edited after redoing composite figures
%111314 ALH few edits
% 111914 NJE moved table call and description here.

\appendix\label{appendix}

\section{Results on Sources}

The full list of sources with \hcop\ observations is given in 
Table \ref{misfitsdat}, along with their spectral index, $\alpha$, 
after extinction correction, as described in
%Dunham
\citet{2013AJ....145...94D},
and the resulting SED class.

\section{Notes on Clouds}

\subsection{Aquila}

Many spectra had two velocity components which could be separated
in some positions, but not all. Emission also moved
in velocity over a range of 3 to 10 \kms.
The two components are not apparent in the average spectrum in
figure \ref{spec2}; instead they blend into a single, rather broad
(3.0 \kms) component (Table \ref{tbl-2}).
The Spitzer data on this cloud were analzyed by 
%Gutermuth et al.
\citet{2008ApJ...673L.151G},
and a preliminary analysis of  the {\it Herschel} data can be found in
%Bontemps et al.
\citet{2010A&A...518L..85B}.

\subsection{Auriga}

This cloud is also called the California Molecular Cloud.
There are at least two velocity components, one near $-1$ \kms\ and another
near $-4.5$ \kms, along with a possible third comoponent around 1.5 \kms. 
Only the component centered near $-1$ \kms\ is strong
and fitted in the average spectrum in figure \ref{spec3}.
For the most part, each position shows one component, but
in a few positions, both were present and two Gaussians
were fitted. 
The Spitzer data were published by 
%Broekhoven-Fiene et al. preprint
\citet{2014ApJ...786...37B} and Herschel imaging by 
% Harvey et al. Herschel and CSO
\citet{2013ApJ...764..133H}.
A single source was observed in Auriga North, and its spectrum 
in figure \ref{spec3} shows that it has a quite different velocity
$-6.9$ \kms\ from the main Auriga cloud.

\subsection{Cepheus clouds}

Most of these clouds tended to have a single velocity component in the
range of 1-5 \kms. Cepheus 1 had two components in some positions,
both in the range of $-5$ to $-3$ \kms, but these could also be self-absorbed
spectra from a single component. We have fitted the average spectrum
for Cepheus 1 with two components (Fig. \ref{spec3}). The average spectra 
of the other two clouds with detections
(Fig. \ref{spec3} for Cepheus 5; Fig. \ref{spec1} for Cepheus
3) show relatively narrow lines, and they were fitted by a single
Gaussian  (Table \ref{tbl-2}. 
Cepheus 5 had a line centered at $-7.8$ \kms, while the line in Cepheus 3
is centered at $+2.5$ \kms.  
The Spitzer data for these clouds have been analyzed by
%Kirk et al. (2009)
\citet{2009ApJS..185..198K}.
The YSOs in Cepheus 1 are associated with L1251A and L1251B.
These regions have been analyzed in detail  
% Lee et al. 2006, 2007 (L1251B) and 2010 (L1251A)
(\citealt{2006ApJ...648..491L,2007ApJ...671.1748L,2010ApJ...709L..74L})
showing that the \hcop\ \jj32\ spectrum is self-reversed toward the
collection of YSOs, so the two components in our average spectrum
are most likely not separate components but instead manifestations of
a self-reversed profile.

\subsection{Chamaeleon Clouds}

The Cha I cloud showed two possible velocity features, but they were
not very apparent in the average spectrum, which was not well fitted by a single
Gaussian (Fig. \ref{spec1}).  However attempts to fit two Gaussians did not
produce believable results.
The components were always close together and could be simply self
reversed profiles of a single component.
The Cha II cloud had a single narrow component at the two positions with
detections. The average spectrum is in figure \ref{spec3}.
The Spitzer data for Cha II were published by 
% Alcala et al. 2008
\citet{2008ApJ...676..427A}
and Herschel data on Cha II have been analyzed by 
% Spezzi et al.
\citet{2013A&A...555A..71S}.

\subsection{Corona Australis}

This cloud exhibits one primary velocity component, 
but some lines were self-absorbed and/or showed broad wings.
The average spectrum (Fig. \ref{spec1}) shows some structure, but
only a single component could be easily fitted.
The {\it Spitzer} data on this cloud have been published by
% Peterson et al.
\citet{2011ApJS..194...43P},
and a detailed analysis of the main sources with data from {\it Herschel}
is presented by 
% Lindberg 2013 DIGIT paper
\citet{2014A&A...565A..29L}.

\subsection{IC5146}

This region has two cloud sections, IC5146-NW and IC5146-E.
IC5146-NW has a single velocity component centered around  4 \kms. 
The average spectrum (Fig. \ref{spec2}) was fitted
with a single component, although there is clearly some non-Gaussian structure.
The average spectrum for IC5146-E was fitted with two components, one
at 6.5 \kms, and stronger one at 8 \kms. 
The Spitzer data have been published by
% Harvey et al. 2008
\citet{2008ApJ...680..495H}.

\subsection{Lupus}

The Lupus region contains 6 clouds, usually denoted by roman numerals
\citep{2008hsf2.book..295C}. The only source detected in \hcop\ was
in Lupus III (Fig. \ref{spec1}). The velocity from that detection was
used to set upper limits on the other sources. The Spitzer data were
published by
%Merin et al.
\citet{2008ApJS..177..551M}.

\subsection{Musca}

No line was detected toward the Musca cloud, so the upper limits
are based on the velocity range of \coo\ enission in 
%Vilas-Boas (1994), ApJ 433, 96
\citet{1994ApJ...433...96V}.

\subsection{Ophiuchus}

Many spectra had two velocity components which could be separated
in some positions, but not all. Both components also moved
in velocity over a range of 2.5 to 5 \kms. 
The average spectrum (Fig. \ref{spec2}) has been fitted with
two components (Table \ref{tbl-2}), and there is also a shoulder on the
high-velocity side of the spectrum.
This cloud was mapped in CO and \coo\ by 
%Ridge et al. 2006
\citet{2006AJ....131.2921R},
who provided fits to their cloud averaged spectra.
The mean velocity for both species was $3.38$ \kms, with linewidths
of $2.77$ \kms\ for CO and $2.38$ \kms\ for \coo.
Both linewidths are substantially larger than those of our
individual components and may represent blends. Their average
spectra are consistent with our two components, but they were too
blended to fit separately.

\subsection{Perseus}

Two components are seen in the spectra of many positions. They are
quite blended in the average spectrum (Fig. \ref{spec2}, but two
components could be fitted. This spectrum is broadly consistent with
an average spectrum of \coo\ in 
%Ridge et al. 2006
\citet{2006AJ....131.2921R},
which is however more blended. Their spectrum shows a third peak 
around 0 \kms, which may appear weakly in our average spectrum.
\citet{2006AJ....131.2921R} also note a velocity gradient from east to
west across the cloud.

\subsection{Serpens}

Many spectra had two velocity components (7.6 and 9.6 \kms), 
which could be separated
in some positions, but not all. Both components also moved
in velocity over a range of 7 to 8 and 8 to 10 \kms, respectively.
The average spectrum (Fig. \ref{spec2}) was fitted with two components
(Table \ref{tbl-2}, and there is weak shoulder extending to higher velocity.

%long MISFITS properties table:
\clearpage
\LongTables
\begin{landscape}
% Table 5  10/15/14 ALH
%111314 ALH added velocities, corrections
% 020415 NJE minor edits
%\begin{landscape}

\end{deluxetable}
%\label{table5}
%\clearpage
%\end{landscape}

\clearpage
\end{landscape}

\clearpage

\end{document}